\documentstyle[aasms4]{article}

\received{}
\accepted{}
\journalid{000}{15 February 1999}

\begin{document}

{

\title{Deep Intermediate-Band Surface Photometry of NGC~5907}

\author{
Zhongyuan Zheng\altaffilmark{1,3}, 
Zhaohui Shang\altaffilmark{1,2,3},
Hongjun Su\altaffilmark{1},
David Burstein\altaffilmark{5}, 
Jiansheng Chen\altaffilmark{1,3},
Zugan Deng\altaffilmark{6,3}, 
Yong-Ik Byun\altaffilmark{4,9}, 
Rui Chen\altaffilmark{1,3}, 
Wen-ping Chen\altaffilmark{4},
Licai Deng\altaffilmark{1,3},
Xiaohui Fan\altaffilmark{1,7}, 
Li Zhi Fang\altaffilmark{10}
Jeff J. Hester\altaffilmark{5},
Zhaoji Jiang\altaffilmark{1,3}, 
Yong Li\altaffilmark{5},
Weipeng Lin\altaffilmark{1,3}, 
Wei-hsin Sun\altaffilmark{4},
Wean-shun Tsay\altaffilmark{4}, 
Rogier A. Windhorst\altaffilmark{5},
Hong Wu\altaffilmark{1,3}, 
Xiaoyang Xia\altaffilmark{8,3},
Wen Xu\altaffilmark{1,5},  
Suijian Xue\altaffilmark{1,3},
Haojing Yan\altaffilmark{1,5}, 
Zheng Zheng\altaffilmark{1,3},
Xu Zhou\altaffilmark{1,3}, 
Jin Zhu\altaffilmark{1,3} and 
Zhenglong Zou\altaffilmark{1,3}
}

\altaffiltext{1}{Beijing Astronomical Observatory, Chinese Academy of
    Sciences, Beijing 100080, P. R. China}
\altaffiltext{2}{Department of Astronomy, University of Texas at Austin,
    Austin, TX 78712}
\altaffiltext{3} {Chinese Academy of Sciences--Peking University joint
    Beijing Astrophysics Center, Beijing, 100871, P. R. China}
\altaffiltext{4}{Institute of Astronomy, National Central University,
    Chung-Li, Taiwan}
\altaffiltext{5}{Department of Physics and Astronomy, Box 871504, Arizona
    State University, Tempe, AZ  85287--1504}
\altaffiltext{6}{Graduate School, Chinese Academy of Sciences, Beijing
    100080, P. R. China}
\altaffiltext{7}{Princeton University Observatory, Princeton, New
    Jersey, 08544}
\altaffiltext{8}{Department of Physics, Tianjin Normal University, P. R. China}
\altaffiltext{9}{Center for Space Astrophysics and Department of Astronomy,
Yonsei University, Seoul, 120--749, Korea}
\altaffiltext{10}{Department of Physics, University of Arizona, Tucson, Arizona,
85721}

\begin{abstract}

Intrigued by the initial report of an extended lumiosity distribution
perpendicular to the disk of the edge-on Sc galaxy NGC~5907, we have obtained
very deep exposures of this galaxy with a Schmidt telescope, large-format CCD,
and intermediate-band filters centered at 6660{\AA} and 8020{\AA}.  These two
filters, part of a 15-filter set, are custom-designed to avoid the brightest
(and most variable) night sky lines. As a result, our images are able to go
deeper, with lower sky noise than those taken with broad-band filters at
similar effective wavelengths: e.g., 0.6 e$^-$ arcsec$^{-2}$ sec$^{-1}$ for
our observations vs. 7.4 e$^-$ arcsec$^{-2}$ sec$^{-1}$ for the R-band
measures of Morrison et al. In our assessment of both random and systematic
errors, we show that the flux level where the errors of observation reach 1
mag arcsec$^{-2}$ are 29.00 mag arcsec$^{-2}$ in the 6660$\rm \AA$ image
(corresponding to 28.7 in R-band) and 27.4 mag arcsec$^{-2}$ in the 8020$\rm
\AA$ image (essentially on the I-band system).

As detailed in Shang et al., our observations show NGC~5907 has a luminous
ring around it that most plausibly is due to the tidal disruption of a dwarf
spheroidal galaxy by the much more massive spiral.  Here we show that,
fainter than 27th R mag arcsec$^{-2}$, the surface brightness around NGC~5907
is strongly asymmetric, being mostly brighter on NW (ring) side of the galaxy
midplane.  This asymmetry rules out a halo for the origin of the faint
surface brightness we see.  We find this asymmetry is likely an artifact
owing to a combination of ring light and residual surface brightness at faint
levels from stars that our star-masking procedure cannot completely eliminate.
The possible existence of an optical face-on warp in NGC~5907, suggested by
our VLA HI observations, is too confused with foreground star contamination
to be independently studied.

Good agreement with the surface photometry of NGC~5907 by Morrison et al. 
and other workers lead us to conclude that their data are similarly
affected at faint levels by ring light and residual effects from their star 
masking procedures.  Inspection of the images published by Morrison et al. and 
Sackett et al. confirm this to be the case.  Thus, we conclude that NGC~5907 
does not have a faint, extended halo.

\end{abstract}

\keywords{galaxies: individual (NGC~5907) -- galaxies: interactions -- 
galaxies: spiral -- galaxies: photometry}

\section{INTRODUCTION}

Until \cite{Sackett94} and Morrison, Boroson \& Harding (1994; hereafter MBH)
announced discovery of a faint, luminous light distribution around NGC~5907,
this galaxy was primarily known as a typical, edge-on Sc galaxy which happens
to be near enough to us ($\rm V_{\rm helio} = 667$ km s$^{-1}$ = 11 Mpc for
$\rm H_0 = 65$ km s$^{-1}$ Mpc$^{-1}$) that we can study it in some detail. 
Previous observations by several groups (e.g. \cite{Sancisi76}; \cite{van81};
\cite{Skurtskie85}; \cite{Sasaki87}; \cite{Barnaby92}) had shown this galaxy
to have both an HI and optical warp, but no obvious galaxy companions to
produce this warp.  As such, NGC~5907 became the prototype of the class of
spiral galaxies having ``non-interacting warps'' (cf. \cite{Sancisi76}).

The observations of \cite{Sackett94} sparked new interest in this galaxy, in
that they claimed to detect a significant halo around it.  Two other
groups then obtained deep surface photometry to try to study the halo in
various passbands (\cite{Lequeux96}, \cite{Lequeux98}; Rudy et al. 1997).  Our
collaboration (the Beijing-Arizona-Taipei-Connecticut (BATC) Color Survey of
the Sky (\cite{Fan96}) already had NGC~5907 as part of our galaxy calibration
program.  The BATC program uses the 0.6/0.9m Schmidt telescope at the Xinglong
Station of Beijing Astronomical Observatory (the ``BAO Schmidt''), with its
focal plane equipped with a $2048 \times 2048$ Ford CCD. We have
custom-designed a set of 15 intermediate-band filters to do spectrophotometry
for pre-selected 1 deg$^2$ regions of the northern sky with this CCD system
(cf. \cite{Fan96}).

\cite{Shang98} summarize the main results of both our deep surface photometry
and our Very Large Array\footnote{The Very Large Array of the National
Radio Astronomy Observatory (NRAO) is operated by Associated Universities,
Inc., under cooperative agreement with the National Science Foundation.}
(VLA) 21 cm HI observations of NGC~5907:  a) We detect a faint luminous ring 
around the galaxy, plausibly due to the recent tidal disruption of a nearby 
dwarf spheroidal galaxy.  b) The HI map picked up a companion dwarf irregular 
galaxy, PGC~54419, on the side of one of the HI--warps and separated in 
radial velocity by only 45 km s$^{-1}$ from NGC~5907.  As such, we move 
NGC~5907 out of the category of ``non-interacting'' warped galaxies, 
but whether the observed interactions are sufficient to produce the observed 
warp is left for galaxy modelers to decide.  c) Our HI observations suggest
that the HI layer is higher near the center of the galaxy than at larger
radii, suggesting this galaxy may also have a face-on warping of the disk.

In this paper we present the details of our deep surface photometry of this
galaxy, this time paying close attention to the issue of the faint luminosity 
distribution around this galaxy.  Section 2 presents our observations,
including details of the data reduction process, which are important for the
reader to be able to critically assess the accuracy of our method.  In Section
3 we study the faint luminosity distribution around NGC~5907 as it appears in
our images, including how the ring, foreground stars, and a possible face-on 
warp can influence what we see.  Our results are compared to those previously 
published in Section 4, in which we also reassess the likelihood of a halo
existing around this galaxy.  Section 5 summarizes the main results of this 
paper.

\section{OBSERVATIONS AND DATA REDUCTION}

The phases of data reduction that contribute to errors in the final image and
over which we have some control are: bias subtraction, dark subtraction,
flatfielding, sky background-fitting, star-masking, and photometric
calibration. In this section we discuss these data reduction issues as they
apply to the data we have obtained.  When we compare our surface photometry 
to those of other observers (Section 4), we will find that, with few 
exceptions, our data are in very good agreement with published results.

\subsection{Data Log and Preselection of Good Images}

The observations of NGC5907 were obtained with the BAO 0.6/0.9m Schmidt
telescope, using a thick Ford 2048$\times$2048 CCD having 15$\mu$m pixels at
the f/3 prime focus. The field of view of this CCD is $58'$$\times$$58'$ and
the scale is $1.71''$ per pixel.  With this combination of telescope and CCD
system, we can comfortably observe large extended objects and also get a
sufficient amount of sky in a single frame.  A large pixel size is better
for finding low surface brightness features, as it minimizes the number of
CCD pixels per surface area.  

The images were taken with slight shifts between exposures (``dithered'') so
that cosmic rays and defects on CCD could be removed during the combining
process.  We employ the Lick Observatory data-taking system, which
automatically subtracts the overscan of each image at readout time
(``on--the--fly''), and records the overscan in the last column. We then
process each program image through a series of software routines we term
PIPELINE--I, which, among other operations, median-filters the original
overscan and adds back to the image the difference between the original
subtracted overscan and this filtered overscan. In this way, the
signal-to-noise (S/N) of the overscan is increased and any residual
large-scale pattern produced in the image by imperfections in the overscan are
removed.

Exposures of the field around NGC~5907 were made through two BATC filters:
6660$\rm \AA$ ($\rm m_{6660}$) and 8020$\rm \AA$ ($\rm m_{8020}$), whose band
widths are 480$\rm \AA$ and 260$\rm \AA$ respectively (cf. Figure~1 in
\cite{Fan96}).  The BATC filters are designed to avoid contamination by the
brightest and most variable night sky emission lines, producing a sky
background of lower surface brightness obtainable under the same conditions in
the broad band R and I filters. Use of intermediate bands also minimizes
effective wavelength-related effects in the CCD sensitivity (Section 4.4).

137 frames of either 10m or 20m duration were obtained with the 6660$\rm \AA$
filter on 23 nights from January 31, 1995 to June 27, 1995, all during
moonless intervals at airmasses $\le 1.5$. The CCD was run with a gain of 4.1
e$^-$/ADU and a readout noise of 12 e$^-$. Similarly, 63 frames of 20m
duration were obtained similarly with the 8020$\rm \AA$ filter on 11 nights
from May 26, 1995 to April 18, 1996, during which time the readout noise
remained as before, but the gain varied somewhat.  For the final combined
8020$\rm \AA$ image the equivalent gain is 3.9 e$^-$/ADU.  Of these 200
images, we selected the images for final reduction in the following manner
(cf. Tables 1 and 2):

(1) Each individual frame was inspected visually to see if there are abnormal
events that might affect photometry, e.g., traces of satellites crossing
galaxy area, or visible problems with the bias overscan. If there was anything 
abnormal, the image was rejected.

(2) For each frame we measured the mean sky level, FWHM and flux for 20
isolated stars found in common on all frames.  These values are shown in
Figure~1 (and tabulated in Table 1) as a function of sequential image number
for the 137 6660$\rm \AA$ images, with flux and sky level normalized to an
integration time of 20m.  Those images having high stellar flux levels, low
sky levels and reasonable seeing were selected for further study. A similar
analysis was done for the 8020$\rm \AA$ images (Table 2), with the net result
that the final combined images are made of 84 6660$\rm \AA$ frames taken on
15 nights and 50 8020$\rm \AA$ frames taken on 8 nights. The total exposure
time in the combined images is $\rm 26^h10^m$ (94200s) and $\rm 16^h40^m$
(60000s) respectively. The images which are used are listed in Tables 1 and 2
with their sky level counts, FWHM of stars and the mean counts (for 20m
integration) measured for the 20 comparison stars.

\subsection{Bias and Dark Subtraction}

Biases measure the ADU level of zero second exposures and, as such are of low
S/N individually.  We took an average of 26 bias frames per night of
observation using the IRAF task ``imcombine'', choosing the minmax algorithm
in combination with the parameters ``nlow'' and ``nhigh'' set to 3 to reject
bad points at $\ge 3 \sigma$ level.  As one test of CCD stability, we take 30 
bias averages from a two month period during the observations and divide them 
into two sets of 15 bias averages each: BIAS1 from the first 15 bias averages;
BIAS2 from the second 15. The difference, BIAS1--BIAS2 has a mean value of
zero and no large-scale pattern exists, indicating good stability of our
data-taking system.  The average of BIAS1 and BIAS2, hereafter BIAS, is
adopted as the final bias frame for correcting. It is equal to the average of
600 individual bias frames, has a mean value of about 1.2 ADU and a standard
error of about 1 ADU/pixel (cf. Sec. 2.7).

Many dark frames must also be obtained throughout the observation sequence to
reliably measure the dark current of the CCD. The dark count was obtained by
obtaining forty, 20m dark frames over the same two month period. Subtracting
BIAS from these frames yields a very small dark count of $\sim 0.25$
ADU/pixel/20m exposure.  This summed dark count frame was then rescaled to the
exposure time of, and subtracted from, each program image.

\subsection{Large-scale Homogeneity of the Program Image}

The first requirement in obtaining homogeneity over large-scales in a CCD image
is to produce a very accurate flatfield map of the large-scale variations that
exist in CCDs, and for those large-scale variations to be stable over
periods of time long compared to the time over which the flatfield images 
and program images are obtained.  This, in turn, requires controlling two
factors:  a) the DC stability of the CCD itself over the image area of
interest; and b) the flatness of night sky and twilight, which, of course, are
never really flat over $>10'$ angles (see below).  

A third factor affecting how well one can subtract the sky from the image of a 
galaxy is the fact that the quantum efficiency of a CCD pixel is a function 
of the effective wavelength of the radiation it detects.  The one part of the 
data-taking process we {\it cannot} control is that the galaxies have 
different spectral energy distributions (SEDs) within the filters, and that 
these SEDs are different from that of the night sky.

While we cannot completely eliminate the systematic effects on the large
scale homogeneity of any of these issues, we have taken steps to minimize
their influence:

{\bf a. DC--stability:}  We can comfortably fit a galaxy as large as NGC~5907 
(14$'$ in diameter) within the inner 25\% of the CCD, which itself is in the 
inner 20\% of the field of a Schmidt telescope.  This eliminates issues 
pertaining to edge-effects in CCD flatfields and minimizes non-uniformity 
issues related to the optics of the telescope. 

{\bf b. Flatness of the Flatfields:} On a CCD chip covering essentially
$1^\circ$$\times$$1^\circ$ of the sky, neither the twilight nor the night sky
will be flat over the whole chip (\cite{Wild97}).  This is obvious upon visual
inspection of the twilight sky, when one takes into account that the field of
view of our CCD is twice the size of the full Moon.  It is also true on a
cloudless, moonless night, as the airglow brightness correlates with air mass.
Our eyes can easily detect gradients in twilight the size of the full Moon,
and at a dark site, gradients in the sky over degree-sized scales. One might
be able to devise a methodology of taking night sky flats at various airmasses
and sky directions to minimize such gradients in the night sky, but such
techniques are likely to be very time consuming, with no guarantee that sky
gradients can be completely eliminated.

Rather, we take advantage of the special optical properties of a Schmidt 
telescope to construct a reliable flatfielding method that can employ a dome 
light.  As detailed in \cite{Chen99}, we place a UV-transparent plastic 
diffuser over the correcting lens of the Schmidt.  One dome light flatfield 
frame has more than 20000 ADU per pixel, corresponding to more than 84000 
electrons (gain = 4.1).  Twelve dome flats produce an overall flat field whose 
statistical pixel-to-pixel error is less than 0.1\%.  The diffuser 
ensures that the flux entering the Schmidt telescope is of a highly uniform 
nature.  In the present paper, the proof of the accuracy of this flatfielding
technique is demonstrated in two ways: by comparing noise in the division
of average flatfields taken on successive nights to that expected from photon 
noise alone, and by the accuracy to which we can flatten the sky.

{\bf c. Color-related Effects:}  One of the factors leading us to decide
to use intermediate-band filters for the BATC survey is the fact that 
SED-related quantum efficiency effects do exist in CCDs.  Our own tests with 
stars (cf. \cite{Fan96}) show the measurements made in the two filters 
used here are insensitive to color changes in the program objects at
the 1\% level.  Such is not necessarily the case for observations made with
broad-band filters (\cite{Stetson90}).

\subsection{Rectification of the Summed Image}

After the correction for the flatfield is applied, position offsets among the
selected CCD frames were calculated with six plate coefficients and the
position of the frame center, using the Guide Star Catalog of the Space
Telescope Science Institute (\cite{Lasker90}).  This process is done
during the PIPELINE--I reduction process, and the derived plate coefficients 
are put into the FITS image header. After shifting all corrected images to a 
common center, bad pixels and cosmic-rays are rejected, and the cleaned images
are combined to a single frame. We then redetermine the plate coefficients and
plate center for the combined frames. The net images in the 6660$\rm \AA$ and
8020$\rm \AA$ filters contain 1928$\times$1969 pixels and 1979$\times$1979
pixels respectively. Figure~2a shows the combined image in the 6660$\rm \AA$
filter, Figure~2b shows the combined image in the 8020$\rm \AA$ filter.

\subsection{Prediction of Background under the Galaxy and Ring} 

A reliable estimate of the surface brightness at faint levels near NGC~5907
requires us to be able to reliably determine the sky background under the 
regions of interest.  These include not only the regions immediately adjacent
to NGC~5907, but also for the ring which we found around it.  This, in turn, 
means modeling the background formed not only by the night sky, but also by 
extraneous sources of light, including stars, other galaxies and, if present, 
Galactic ``cirrus.''

\subsubsection{Star and Galaxy Masking}

The stars and background galaxies to be subtracted on the combined image range
from faint, near-point-sources of light to large regions contaminated by
highly saturated stars.  The traditional way to handle these sources of extra
light is to mask them until their wings fall under a certain level, say, 10\%
of the variation in the sky.  This is the procedure followed by MBH and 
Lequeux et al. (1996). We found that if we followed this procedure with our 
own combined images, we would be left with too few sky pixels from which to 
determine a reliable sky background.  As a result, we proceed in a different, 
stepwise manner.

First, using DAOPHOT (\cite{Stetson87}), we fit a point-spread-function
(PSF) to each combined image.  The fitted PSF is then subtracted from star
images that are not saturated, following the prescription given in the 1987
Stetson paper and in the DAOPHOTII User's Manual.  Saturated stars were not
subtracted.  Key to this process is as accurate a subtraction of the wings of 
the PSF as one can practically make.  

The next step is to check the reliability of the PSF subtraction by plotting 
the residuals (obs$-$PSF model) versus distance from PSF center for each 
unsaturated star.  The PSF--wings are subtracted cleanly for most stars and 
there is little remnant.  However, the central areas of these 
subtracted stars show higher statistical fluctuations than the wings, owing
to the higher original signal in the centers.  While circular masks are
placed over such cases, the size of the circles cover far less area than 
would have been the case if the whole PSF--wings of every star was masked.

The brightest, unsaturated stars are the exception to this second step, as the
PSF-subtraction leaves a noticeable residual at faint light levels.  Therefore,
we treat both the brightest stars, saturated and unsaturated, the same,
masking out the whole star to faint, but finite sky levels. In total, 117 
bright stars in the 6660{\AA} image were masked in this way.

Other galaxies that are resolved on our images are easily found via the DAOPHOT
star-subtraction process, as the PSF-subtraction leaves a deficit of flux at
the galaxy center surrounded by a bright halo.  All background galaxies are
masked entirely, but are generally so faint that these mask radii are small.

Figure~3 shows four results of the PSF-subtraction, including a saturated star,
a bright unsaturated star fit by the PSF, a faint star fit by the PSF and for
a faint galaxy fit by the PSF.  The vertical lines indicate masking radii
used.  Our tests on all PSF-subtracted stars show that the residuals of the
PSF-fits are less than 5 ADU on average, or approximately 5\% of the
statistical {\it noise} of the sky.  

Treating saturated stars in these images is more problematic.  The images of
star which cannot be PSF-fitted due to saturation effects should, in
principle, be wholly masked.  Unfortunately, as stated earlier, if we tried to
completely mask every bright star in our image, we would not be left with many
pixels left in the image!  As a result, we are forced to make two compromises
in our star-masking procedure.  First, we only mask each bright star to a
surface brightness of 29th mag/arcsec$^2$.  Second, owing to the optics of a
Schmidt telescope, the PSF's of the stars are not symmetric anywhere in focal
plane of our telescope.  This is easily seen in Figures 2a and 2b of the
present paper. However, the masks are, by necessity, symmetric. The net result
is a slight mismatch of mask and PSF at very low light levels. As we will see,
these compromises limit what we can do at faint light levels around NGC~5907.

In our 6660$\rm \AA$ combined image, there are 7921 unsaturated objects
detected among which are 1694 background galaxies in 1000$\times$1000 pixel
(1710$''$$\times$1710$''$). We also note that, if one compares our 6660$\rm
\AA$ combined image with Figure~2 of MBH and Figure~1 of Sackett et al., it
is clear that our observation is as deep, if not deeper, than theirs,
detecting all of their stars, plus some fainter stars.  Quantiatively, our
detection threshold is 3.5$\rm \sigma_{\rm sky}$, resulting in a limited
magnitude of 23.3. This means that for $\rm m_{6660} \leq 23.30$, $\rm
\log(N_{\rm star}) = 4.45$ deg$^{-2}$ and $\rm log(N_{\rm galaxy}) = 3.88$
deg$^{-2}$. Star counts are overestimated, and galaxy counts are underestimated,
as galaxies that are unresolved appear stellar, and galaxies with too low
surface brightnesses are not found.

Separately, we mask out a circle of radius 250 pixels around the galaxy, as
well as an area around the ring that is otherwise not excluded.  This ensures
that no galaxy-related or ring-related flux are included in the sky background
determination, but it also means that the residual flux from bright stars
masked in these regions is not modeled.  The final masked 6660{\AA} image is 
shown in Figure~4.

\subsubsection{Sky Background Fitting}

Once star-masking and galaxy-masking is complete, we can then attempt to model
the sky background.  Since it only makes sense to do this calculation near the
galaxy and ring, we can safely limit ourselves to the inner 1000 $\times$ 1000
pixel areas ($28.5' \times 28.5'$) in each combined image, the region
marked by the square in Figure~2a.  This is also the region defined by our
masking procedure (Figure~4).

First, we produced a smoothed version of each masked combined image by mode 
filtering each image with a box 10$\times$10 pixels in size. To avoid 
contamination by masked pixels, only those pixels not in a masked area
were used in calculating the mode in any box.  This necessarily means that
near each masked region, fewer pixels were used to generate the mode.
After smoothing the mask regions whose radii are larger than box size are 
shortened and those whose radii are smaller than box size disappear. 
Once this smoothing operation is done, the original mask is reimposed on
the image, thus eliminating the pixels which are in original masked areas.
As a result of this process, edge effects from the masking procedure are 
removed. 

We next experimented with the IRAF task imsurfit to perform fitting the sky
background with various two-dimensional analytic functions. None of these
models was found to be wholly satisfactory, owing to the fact that we have to
restrict the fits to low order polynomials to avoid introducing spurious
interpolations in the masked regions. In every case, we could see large areas
of the image that were systematically underfit or overfit.  Low order spline
fits similarly suffer.  At the end, we settled on a stepwise method that 
works reasonably well when faced with a large part of the image masked-out.

We fit each row of the smoothed-and-masked image with a one-dimensional
Legendre polynomial of low order.  In doing this fit, we reject points above
$2 \sigma$ on the high side and $3 \sigma$ on the low side, as well as all
pixels inside the mask.  The reason for the asymmetric rejection of points is
that the main sources of scatter on the high side are faint, undetected
sources and unfitted faint wings of stars, while the low side values result
from statistical fluctuations.  This is clearly seen when you make histograms
of the pixel values in the masked images; the histograms are skewed towards
positive values.  Necessarily, this process requires interpolating each line 
under the galaxy/ring mask, keeping the order of the Legendre polynomial to 
be 3 or less.

The row-fit to the smoothed 6660$\rm \AA$ image is shown in Figure~5a. Next,
we repeat this process in the column direction (Figure~5b). This
ensures we are predicting the sky underneath the galaxy in a mutually
orthogonal manner. The row and column fits are then averaged (Figure~5c), and
this average image is smoothed with a circular Gaussian of $\sigma = 30$
pixels, truncated at $\pm 4 \sigma$. The final smoothed image was adopted as the
sky background (Figure~5d).

The 6660$\rm \AA$ image with its fitted background subtracted is shown in
Figure~6a and the 8020$\rm \AA$ image is shown in Figure~6b. As is evident,
even with our best attempt at star-subtraction and masking, one can still see
faint, partial rings near many of the brighter stars.  These rings are at
surface brightnesses 29th mag arcsec$^{-2}$ and fainter in the 6660$\rm \AA$
image.  As stated earlier, this end result is a compromise, for if we were to
truly exclude all starlight from this image, very little would be left of the
image from which to determine sky, and in many of the MBH cuts, leaving little
or net flux to examine.  As is plain, NGC~5907 is at a Galactic latitude
(51$^\circ$) where stellar crowding starts to seriously affect what one can
quantitatively do at low surface brightnesses.  As a check of the accuracy of
our background fitting and masking procedures, we have taken four averages of
30 rows (``slices'') each through the background-subtracted 6660$\rm \AA$
image (Figure~7) and similarly-processed 8020$\rm \AA$ image (Figure~8) that
span the region of interest around NGC~5907.

At the tops of each panel of Figures~7 and~8 are given these sky-subtracted
slices through the galaxy.  At the bottom of each panel are given the
residuals of the subtracted sky background for each slice. The upper line is
the residual of our sky fitting and subtracting procedure, while the lower
line is the residual of normal two-dimensional function fitting. They are
shifted down by 110 ADU and 150 ADU seperately for display purposes. It is
obvious that our procedure produces smaller residuals than two-dimensional
function fitting.  From Figures~7 and 8 a reader of this paper can check as 
to whether our sky subtraction procedure introduces any spurious features into 
the sky-subtracted image.  The lower parts of these panels also demonstrate 
that our sky-subtraction procedure conservatively fits an effective 
two-dimensional plane under the masked region around the galaxy.

Of course, no sky background fit is perfect, and it is these imperfections
that contribute to the limit to which one can do accurate faint surface
photometry. To estimate the systematic error from this source, we sample 381
areas of 50$\times$50 pixels of the background-subtracted 6660$\rm \AA$ image
and calculate the mode value of each area.  After excluding areas in the region
severely affected by masking, the standard error of these mode values is 18
ADU.  We take this as our best estimate of the irreducible $1 \sigma$ accuracy
of our sky background fit.  In a similar manner we determined the accuracy of
the sky background fit to be 17 ADU for the 8020$\rm \AA$ image.

In contrast, if we had limited our sky background fits to only those obtained
by analytical 2-dimensional functions, the row slices showed many more
systematic variations and the $1 \sigma$ systematic error was typically 30 ADU. 
We believe our method of sky background fitting produces a reliable
match to what is, inherently, a very lumpy sky at very low light levels.

\subsection{Zero Point Calibration}

It is standard practice for BATC observations to use \cite{Oke83} standard
stars as our calibration stars when the night is photometric.  The nights of
06March95 and 01April96 in the series taken for NGC~5907 were photometric. The
magnitude zero point for each combined image corresponds to a flux of 1
ADU/second is: $19.08 \pm 0.01$ mag for the 6660$\rm \AA$ image; $17.79
\pm 0.01$ mag for the 8020$\rm \AA$ image.  The sky brightness values are
21.26 mag arcsec$^{-2}$ and 19.91 mag arcsec$^{-2}$, respectively.  The
parameters we derive for these two combined images are given in Table~3. Note
that the FWHM of the PSF of the combined images (2.3 and 2.5 pixels, or 2$''$
in radius), is worse than typical seeing of 2$''$ at the Xinglong Observation
Station of the Beijing Astronomical Observatory, using the Schmidt telescope
(cf. Figure~1).

\subsection{The Estimated Errors we Can Control}

The random error per pixel is given by the random error in photoelectrons, not
in ADU.  In the case of the 6660$\rm \AA$ image, there are 4.1e$^-$/ADU, and
3.9e$^-$/ADU for the 8020$\rm \AA$ image.  Hence, if a pixel has $n$ ADU, this
corresponds to $4.1n$ e$^-$ for the 6660$\rm \AA$ image, and the error is
$\sqrt{4.1n}$ in e$^-$, equivalent to an error of $\sqrt{n/4.1}$ in ADU.
In the case of the 8020$\rm \AA$ image, the error would be $\sqrt{n/3.9}$.
The sources of error in our images we can estimate are:  readout noise; photon 
noise per pixel; noise in average bias; noise in average dark; noise in 
flatfielding; systematic errors in the sky background mapping.  Note that all 
of these noise levels save two (systematic flatfielding errors and systematic 
sky background errors) can be reduced by increasing the number of pixels 
sampled.

{\bf Readout noise:}  Random noise from readout comes from a readout noise 
(in terms of error of the mean) of 12 $\rm e^-$/pixel per image.  In the 
case of the 6660$\rm \AA$ combined image (84 input images) this is 
12$\times$$\sqrt{84}\approx$110 $\rm e^-$/pixel; 
12$\times$$\sqrt{50}\approx$85 $\rm e^-$/pixel for the 8020$\rm \AA$
image (50 input images).  From above, the noise in the readout is 
26.8 ADU/pixel for the 6660$\rm \AA$ image and 21.8 ADU/pixel for the 
8020$\rm \AA$ image.

{\bf Photon noise per pixel} is given simply as $\sqrt{n/4.1}$ ADU for the
6660$\rm \AA$ image and $\sqrt{n/3.9}$ for the 8020$\rm \AA$ image.

{\bf Bias and Dark:}
Because the ADU/pixel counts of the average bias and dark are very small, the
main error comes from the readout noise, and is 3 ADU/pixel. The final bias
image used for bias subtraction was an average of 600 single bias frames,
resulting in a noise of 0.12 ADU/pixel. The error transmitted into the
combined image in 6660A is 0.12$\times$$\sqrt{84}\sim$1 ADU/pixel. The
error introduced by dark-subtraction is also about 1 ADU/pixel for both
passbands. Compared to the photon noise of night sky and readout noise of the
program images, the errors introduced by biases and darks are thus negligible.

{\bf Flatfielding}
As is evident from Figures~7~and~8, our flatfielding procedure produces a flat
enough image that the structure in the background is dominated by the
intrinsic lumpiness of the faint night sky.  This is consistent with the tests
we have done to confirm the validity of our flatfielding procedure (Chen et
al. 1999).  However, from the flatfield we do have some pixel-to-pixel noise
due to photon statistics. For each filter, 12 frames of dome flatfields, each
with more than 20000 ADU, are combined together to make the final flatfield
image. Thus, the photon noise in the flatfield is
$\sqrt{20000/4.1}/\sqrt{12}\sim20$ ADU or 0.1\%. Since the final galaxy image
was combined from 84 dithered frames, the error was reduced by 1/$\sqrt{84}$
to 0.01\%, approximately. For the 8020{\AA} image, it is about 0.015\%.

Given the changes in flatfield pattern night-to-night, run-to-run in a typical
CCD system, we separately check the large-scale uniformities of our flatfields
by dividing the combined flat-fielding frame of one night by that of the 
adjacent night.  9 divisions of flatfields taken in the 6660{\AA} band have an 
average pixel-to-pixel variation of 0.16\%, indicating that the noise of the 
combined flat-fielding frame of one night is about 0.11\%.  This value is 
consistent with 0.1\% predicted by calculating photon noise, showing our 
flatfields are very stable from night-to-night.  As such divisions also 
included any large-scale pattern changes in the flatfield, this test also 
shows that the large-scale pattern is reproduced very accurately with the 
dome/diffuser technique for a Schmidt telescope.  When we then smooth the 
flatfield ratios by binning up pixels, we find that the large-scale variation 
is about 0.026\%.  Therefore, the large-scale error of the combined 
flat-fielding frame of one night is slightly smaller than 0.02\%.

{\bf Intrinsic variation in galaxy brightness:}
MBH point out that the intrinsic variation could be expressed by the formula
given by \cite{Tonry88}, assuming $\overline{\rm M}_R=0$ (\cite{Tonry90}). In
our case, $\rm m_1 = 19.08$ mag, an exposure time of 94200s, a distance of 11
Mpc (assuming H$_0$=75). Since the central wavelength of our 6660{\AA} band is
similar to R band and the difference between our 6660{\AA} band and R band is
small (Sec 4.1), a value of $\overline{\rm M}_{6660}=0$ should be a reasonable
assumption. This gives an estimated variance of 3.3g, where g is the mean
number of counts due to the galaxy only. So 45 ADU/pixel (equal to 28.5
mag/arcsec$^2$) will give an error of 12 ADU/pixel, far less than the photon
noise of night sky of 95.1 ADU/pixel. The estimated variance of that in
8020{\AA} image is 1.6g (assuming $\overline{\rm M}_{8020}\sim -1$, when
$\overline{\rm M}_R\sim 0$ and $\overline{\rm M}_I\sim -2$ (\cite{Tonry90})). 
For 22 ADU/pixel (equal to 27.5 mag/arcsec$^2$) we get an error of 6
ADU/pixel, small compared to our other sources of random error.

\subsection{The Total Error Budget}

{\bf Random Noise:}
Of the photon noise of night sky, readout noise, bias/dark count errors and
flatfielding random errors, the reducible random errors in a blank sky area are
dominated by the first, and are estimated to be close to 100 ADU/pixel for the
6660{\AA} image and 83 ADU/pixel for the 8020{\AA} image, which are roughly
consistent with the $\sigma$ measured directly from the combined images (cf.
Table 3). These noises and intrinsic variation of the galaxy can be reduced by
binning adjacent pixels. As the distance from the galaxy plane increases, the
bin size increases exponentially to keep similar S/N ratio. As an example,
the bin size is 45$\times$59 pixels at the farthest distance, and the random
noise is reduced to only 2 ADU.

{\bf Large-scale Error of Flatfielding:} 
Both from the tests we made above, and those made by Chen et al. (1999), the
large-scale error of BATC flatfielding is less than 0.1\% (Chen et al. 1999).
Since we combined 84 frames for 6660{\AA} band and 50 frames for 8020{\AA}
band, the error introduced by this source is reduced to about 0.01\%.

{\bf Systematic Error of Background Subtraction:} In the previous section we
found that the irreducible scatter in the images is 18 ADU for the
combined 6660$\rm \AA$ image and 17 ADU in the combined 8020$\rm \AA$
image. Taken together with the sources of random errors, this places an
intrinsic accuracy of 20 ADU/pixel for faintest signal in the 6660{\AA} image
and 19 ADU/pixel in the 8020{\AA} image {\it where we can explicitly fit the
sky}.  Expressed in terms of surface brightness units, these errors
correspond to a surface brightness of 29.4 mag arcsec$^{-2}$ in the 6660{\AA}
image, and 27.7 mag arcsec$^{-2}$ in the 8020{\AA} image. In masked regions,
this intrinsic error will be larger and more systematic, owing to residual
surface brightness around the masks of bright stars.

{\bf Residual Foreground Star Subtraction:}  The slight imperfections of the 
star masking process can be estimated from the slices in Figures 7 and 8.  
By examination of the regions around the brighter subtracted stars, it is 
apparent that residual star light affects our image at the level of 20--50 
ADU/pixel in specific places around most of the brighter stars.  From above,
this is at surface brightnesses around the 29th mag arcsec$^{-2}$ level 
in the 6660$\rm \AA$ combined image.  It is also evident from Figures 7 and 8
that if we tried to mask the stars to fainter light levels we would, indeed,
be left with very few remaining pixels in these cuts.  The complications
in our analysis of these data that this compromise brings are discussed below.

The sources of variation in our surface photometry in 6660{\AA} that are
generally present in our 6660$\rm \AA$ combined image are summarized
in Table~4. This example assumes we are observing a surface brightness that
gives us 50 ADU counts from the galaxy, 37066 from the sky
and bin size of 45$\times$45 pixels. We choose this bin because it lies at 
the envelope of the galaxy halo.  It can be found that a combination of 
irreducible systematic error of determining sky background, plus unavoidable
imperfect star--masking, determine the ultimate limit of the accuracy that 
can be achieved.  

In reality, to this value (24 ADU) we must add the more problematic residuals 
that we see around bright stars within the masked-out region around the 
galaxy and ring.  At its lowest level ($\sim 20$ ADU), the error is comparable
to our minimum error; at its highest ($\sim 50$ ADU), it dominates the
minimum error.  As we will see, residual surface brightness from bright stars
likely poses the ultimate limit to which we can investigate the halo of
NGC~5907.

\section{SURFACE BRIGHTNESS DISTRIBUTION OF, AND NEAR, NGC~5907}

\subsection{Color and Magnitude of NGC~5907}

We measure the total magnitude of NGC~5907 to a surface brightness of 27 mag
arcsec$^{-2}$ of $\rm m_{6660} = 9.9\pm0.02$ and $\rm m_{8020} = 9.5\pm0.02$.
To obtain these values we replaced the areas excluded due to foreground stars
with the mean value of the surrounding background.  To put this into more
standard terms, we use the fact that we find an average transformation of 
$\rm R-I=m_{6660}-m_{8020}-0.2$ mag, based on comparison with existing surface
photometry of NGC~5907 (Sec. 4.2).  Thus, our observations indicate NGC~5907
itself to have $\rm R-I=0.2\pm0.04$. If we exclude a rectangle around the 
obvious dust lane in this galaxy, we find the average R--I color to be 
$-0.1\pm0.1$.  This bluer color refers to the average value for stars 
less-affected by dust in the disk, and is typical of a stellar population 
dominated by younger stars.

\subsection{The Edge-on Stellar Warp}

The HI warp of NGC5907 first found by Sancisi (1976) is strongly confirmed by
our own VLA 21 cm HI observations (Shang et al. 1998).  Given the warp in the
HI gas, it is logical to search for an analogous warp in the stars. While
\cite{van79} only had marginal evidence for such a stellar warp, Sasaki (1987)
claimed to detect an optical warp at a projected radial distance along the
major axis from 13.3 kpc to 24.0 kpc from the center of the NGC~5907.
\cite{Sancisi93} showed that there are similarities between the optical warp
of Sasaki and the HI warp. In Figure 7 of MBH, there is evidence for the start
of a stellar warp at 4.1 kpc from the center of the galaxy.

The warp in each of our combined images is determined by measuring the
horizontal distances between the surface brightness profiles of two sides of
the galaxy around a fiducial position angle, and measuring their mean offset
as a function of distance along the galaxy major axis. Figure~9 shows the
optical stellar warps calculated from our brightness profiles parallel to the
minor axis.  We find a warp in the galaxy that begins near the center of the
galaxy (at 4.1 kpc, in agreement with what MBH found) and changes continuously
outward to at least 16 kpc from the center, such that at a radius of 16 kpc,
the warp deviates about 0.4 kpc ($\sim$ 4 pixels) above the average galaxy
plane.  We see the same warp in both our 6660$\rm \AA$ image and in our
8020$\rm \AA$ image.  

In regards to investigations at high latitudes for faint surface brightness in
an edge-on galaxy, warps are likely to occur with greater frequency along the
line-of-sight to the disk than at the disk edges, given relative angular
coverage.  Given that a galaxy disk is seen to have a warp at its edges, it
increases the possibility that the galaxy is also warped along the
line-of-sight.  This face-on warp may distort galaxy surface brightnesses at
large z-distances from the galaxy plane in a non-axisymmetric way, as opposed
to the expected symmetry from a halo. As such, separating the two kinds of
possible sources of high-z surface brightness is, in principle, possible. 
Relevant to this issue in the case of NGC~5907, we note that there is some
indication in our current VLA observations, that the HI layer is higher near
the center of the galaxy than outside the center (Shang et al. 1998).  We
explore the consequences of this hypothesis below.

\subsection{The Stellar Ring}

As shown in \cite{Shang98} and has been evident in the deep images we present
here (Figure~6a,b), NGC~5907 has a faint, luminous ring around it. Our
schematic for this ring is shown in the inset in Figure~6a.  As discussed in
\cite{Shang98}, the ring is reasonably elliptical in shape, of similar angular
size as NGC~5907, and the center of NGC~5907 is near one focus of the ellipse.
The reality of this ring has been confirmed on other CCD images (cf. Shang et
al. 1998), and part of the ring can be faintly seen in Figure~3 of MBH
(although they do not recognize it as part of a ring). The various arguments
for the origin of this ring are made in \cite{Shang98}, from which we conclude
that by far the most likely interpretation is that of a dwarf spheroidal
galaxy in the process of being torn apart by a strong tidal encounter with
NGC~5907.

The physical particulars of the ring indicate that the luminosity distribution
within it is irregular both in surface brightness and in apparent thickness.
The maximum width of the ring is found at the NE end of its major axis (40
pixels = 68.4$''$ = 3.6 kpc at the distance of NGC~5907), while its smallest
width is found near the region where the ring overlaps NGC~5907 (20 pixels =
34.2$''$ = 1.8 kpc).  In our estimate of the total brightness of the ring, we
are limited by masked areas within NGC~5907 itself obscuring that part of the
ring nearest the galaxy.  As such, we can only unambiguously measure unmasked
pixels of the ring only in its clearer NE half.  We find the average surface
brightness in the NE half of the ring to be $28.0 \pm 0.3$ mag arcsec$^{-2}$
in the 6660$\rm \AA$ image and $28.3 \pm 1.8$ mag arcsec$^{-2}$ in the
8020$\rm \AA$ image.  The much larger error for the ring surface brightness at
8020$\rm \AA$ is expected, given our error budget estimate (Sec. 2.8).  If we
assume that the whole ring has this average surface brightness, we obtain total
magnitudes of $\rm m_{6660} = 14.7 \pm 0.3$ mag and $\rm m_{8020} = 15.0 \pm
1.8$ for the ring.  However, these estimates may be brighter/fainter limits to
the brightness of the ring, as it is not obvious how irregular is the light
distribution in the regions we cannot investigate due to overlap with the
galaxy and with foreground stars.

The larger error on $\rm m_{8020}$ precludes quoting a reasonable color for
the ring as a whole.  If we measure just the highest surface brightness
features in the ring (denoted by arrows in Figures 8 and 9), we find
26.8$\pm$0.1 mag arcsec$^{-2}$ in 6660$\rm \AA$ and 26.1$\pm$0.2 mag
arcsec$^{-2}$ in 8020$\rm \AA$, yielding a color index $\rm
m_{6660}-m_{8020}=0.7 \pm 0.3$ (or $\rm R-I = 0.5 \pm 0.3$). In order to
confirm this result, we also measure the total flux within the brightest area
of the ring by (indicated by the parallelogram in Figure~6a) to give an
average surface brightness of 27.0$\pm$0.1 mag arcsec$^{-2}$ in 6660$\rm \AA$
and 26.4$\pm$0.3 mag arcsec$^{-2}$ in 8020$\rm \AA$, implying an R--I color of
0.4$\pm$0.4.  Measurement of the second brightest area of the ring also gives
a similar result.  On average, we measure the ring to have an R--I color of
0.5$\pm$0.3, consistent with the R--I colors for Galactic globular clusters of
metallicities [Fe/H] $\sim -1.0$ (cf. \cite{Peterson93}).

\subsection{Cuts Perpendicular to the Disk}

In Figure~6a we show how the perpendicular cuts used by MBH are related to
our combined 6660$\rm \AA$ image.  (Rather than term these as cuts ``along the
minor axis,''  we prefer to designate these as cuts ``perpendicular to the
major axis.'')  As did MBH, we designate each cut by a letter and number, with
A being the central cut, B1 and B2 being the cuts nearest the center (1 to
the northeast, 2 to the southwest), and D1 and D2 being the cuts in the outer
parts of the galaxy.

To increase S/N along these cuts we bin 45 pixels in width and, to keep this
S/N approximately constant, we increase the heights of the bins exponentially
with z distance from the galaxy plane, with 1 pixel being the smallest height,
and 59 pixels being the largest.  This procedure is directly analogous to that
employed by MBH.  For our estimate of galaxy surface brightness, we take the
mode of each bin sampled. The reason for this is that undetected background
sources, plus residuals from the star masking procedure, will tend to upward 
bias the mean and median ADU counts tabulated in each bin.  We experimented 
with alternate definitions of the flux in a bin and found that, while $\sigma$
clipping algorithms can partly resolve this problem, the best way is to use
mode value instead of mean or median.  A $\sigma$ clipping algorithm is used
to eliminate bad pixel values (of which a few still remain in the combined
image) from the histograms.

The perpendicular profiles we derive from the 6660$\rm \AA$ image are given in
Figure~10 and in Figure~11 from the 8020$\rm \AA$ image, and are tabulated in
Table~5 for completeness and to aid in future studies of this galaxy by others.
The profiles of the two sides (NE and SW) about the galaxy plane are folded
around the curved warpline to remove the effect of the edge-on warp. Those
parts of the profiles fainter than $\rm m_{6660} = 29$ and $\rm m_{8020} = 28$
(about 0.1\% of sky) are not presented since their values are uncertain by
over 1 mag owing to the finite size of our limiting systematic errors.  The 
coding of the symbols used to plot these perpendicular profiles is such that 
solid symbols represent parts of the profile {\it not} influenced by the surface
brightness distribution in the ring (Shang et al. 1998), while open symbols
are those part of profiles in which the ring light, and a possible warp in
front of the galaxy, influences what we see.  

It is obvious that at large z-distances from the plane of the galaxy, the
faint surface brightness distribution in the 6660$\rm \AA$ image around
NGC~5907 {\it is not symmetric about the midplane of the galaxy.}  Only in the
case of the A and C2 cuts are the profiles on the NE and SW sides of the
galaxy consistent within the errors at faint surface brightnesses.  The NE
sides of the profiles in cuts C1, D1 and D2 are systematically brighter than
the SW sides in these cuts.  In the case of cuts B1 and B2, one side is
brighter than the other, but the sense of asymmetry changes from B1 to B2. The
B1 profile (to the NW of the galaxy center) has its NE side significantly
brighter than the SW side, while the B2 profile (to the SW of the galaxy
center) has its SW side equally brighter than the NE side as for the B1
profile. Larger errors at faint levels in our 8020$\rm \AA$ image preclude any
such conclusions being separately made for the cuts in this image.

Thus, on the basis of what we see in our 6660$\rm \AA$ image, we find that
the faint surface brightness distribution around NGC~5907 is asymmetric.
With these data, we can confidently rule out the presence of a
halo in this galaxy beyond a distance of 5 kpc.

\subsection{Ring, Foreground Stars and Possible Warp influences on the 
Asymmetry of the Perpendicular Profiles}

As can be seen by comparing the perpendicular profiles in Figures 10 and 11 to
the positions of the cuts relative to the ring in Figure 6a, cuts C1, D1 and
D2 all intersect the ring on the NE side of the galaxy.  What confusion with
the ring can do to the perpendicular profile is also seen in Panel 1 of
Figure~7.  The apparently smooth gradient in the outer profile of the
galaxy (from pixel numbers 600 to 750) is due almost entirely to the 
presence of the ring in this region (cf. Figure~6a).

If we conservatively take the average surface brightness of the ring (28th
mag/arcsec$^{-2}$; in the 6660$\rm \AA$ image; see Sec. 3.3) as our estimate
for the ring surface brightness near the galaxy, then most of the faint
extensions seen in cuts C1, D1 and D2 on the NE side of the galaxy can be
attributed to ring light.  Separately, as we can make a prima facie case
that the ring is elliptical in shape and has the center of NGC~5907 at one of
its foci (cf. Shang et al. 1998), it is difficult to disentangle true galaxy
light from possible ring light in the A cut on the SW side of the galaxy at
surface brightnesses fainter than 28th mag/arcsec$^{-2}$ in the 6660$\rm \AA$
image.

Of the remaining 3 of the six non-centered (B,C,D) cuts made, C2 has the least
foreground star contamination, but still has some star contamination on its SW
side.  In contrast, both B1 and B2 include large sections of the masks around 
bright stars: B2 on its SW side, B1 on its NE side.  Similarly C1 on its NE 
side, D1 on its NE side, and D2 on its SW side.  Interestingly, in every case, 
the side of the galaxy that is brighter is more foreground star-contaminated 
than its partner.  This would strongly suggest that residual star light at 
low light levels (surface brightnesses of 29th 6660$\rm \AA$ mag arcsec$^{-2}$
and fainter) is are contributing to the differential effect in these cuts seen 
at surface brightnesses of 28th 6660$\rm \AA$ mag arcsec$^{-2}$ and fainter. 

On the other hand, {\it if} this galaxy has a face-on warp, one might expect
that the warp would give an asymmetrical distribution perpendicular to the
center of the disk near the galaxy center, i.e., in cut B1 versus cut B2.  
While such an effect is certainly seen, the apparently irreducible effects of 
foreground star contamination make it highly unlikely that we will be able to 
unambiguously detect a face-on warp in this galaxy, even if it exists.
NGC~5907 simply has too many foreground stars around it to be a good edge-on 
galaxy candidate in which to find a face-on warp.

\section{Comparison with Previous Results}

\subsection{Comparison to the Results of MBH.}

In order to do a close one--to--one comparison to the published results of
MBH, Dr. H. Morrison kindly sent us the surface brightness data they used to
make their perpendicular cut diagrams in the MBH paper (data which were not
published in their paper).  We list these data in Table~6 in the same manner
as we did for our own data. If we compare the broad-band R surface brightness
profiles from MBH given in Table~6 to our own 6660$\rm \AA$ surface brightness
profiles given in Table 5, it is apparent that, for similar errors, our data
go $\sim 1.5$ mag fainter than those of MBH. The main reason for this is that
the sky is much fainter in our intermediate band filter (which avoids the
brightest night sky lines) than in the broad-band R filter used by MBH: 0.6
e$^-$ s$^{-1}$ arcsec$^{-2}$ in our combined 6660$\rm \AA$ image to 7.4 e$^-$
s$^{-1}$ arcsec$^{-2}$ quoted by MBH for their R-band data.

As is evident from the graph presented in their paper, Sackett et al.
combined NE and SW sides of each cut to form one combined perpendicular
profile per cut.  With our higher accuracy at fainter light levels, we
can see the asymmetry in the faint light distribution perpendicular to the
disk that MBH and Sackett et al. could not.

We compare the perpendicular profiles of MBH to our 6660$\rm \AA$ image
profiles in Figure~12. A zero point of 0.3 is added to the R band data of MBH
to bring the two filter systems onto the same photometric system at bright
levels of the galaxy.  While we note that MBH cite results only for 
R band surface brightnesses brighter than 27th mag arcsec$^{-2}$, the actual
data given to us by Dr. Morrison go fainter than this.  As neither MBH or
Sackett et al. give actual surface brightness profile information, we choose
to use the data sent to us by Dr. Morrison as the basis for our comparison.

We find generally good agreement with MBH data, even at low light levels, in 
most cases.  Good agreement is found between our data and those of MBH, both in
form and zero point, for distances less than 5 kpc from the galaxy midplane in
all cuts.  This gives strong support to accuracy of the zero points obtained 
by both us and MBH, and to the fact that our perpendicular profiles are 
similar, typically at surface brightnesses brighter than 27th mag arcsec$^{-2}$.

Above a distance of 5 kpc from the galaxy midplane, the comparison must take
into account the asymmetry of the faint light distribution seen. Namely, for
the cuts B1, B2, C1, D1 and D2, the two faintest measurements of MBH are
likely measuring only the brighter flux from one side of the galaxy (else, we
should see the MBH data to be 0.75 mag fainter than ours owing to this
asymmetry).  This is consistent with what we see in Figure~12, in which the
faintest MBH measurements are in accord with our brighter surface
brightnesses measured on one side of the galaxy or the other. The two
exceptions to this agreement at low light levels are the A and C2 cuts, where
the surface brightness difference systematically occurs at levels 27.5 mag
arcsec$^{-2}$ and fainter in these two cuts.

This comparison would strongly suggest that as well as in our data, the
surface brightness profiles determined by MBH are affected by residual
foreground star subtraction at very low light levels.  Inspection of Figure~3
of MBH, and the higher resolution modeled figure in Sackett et al., show this
to be the case.  One can see extra flux around the brighter masked stars in a
similar kind of asymmetric pattern as we see in our own masked 6660$\rm \AA$
image (Figure~6a).  In addition, the two bright stars that are at the lower
left from the galaxy in their image show CCD bleeding along the columns that
is not completely taken out by their masking procedure.

Thus, we conclude that our data, both in zero point and in form, are in good
agreement with the data of MBH at low light levels.  Differences that are seen
are as likely to be attributable to issues pertaining to foreground star
subtraction as to any other kind of systematic error.  From this we conclude
that the data used by MBH and Sackett et al. for their investigations into the
possible thick disk and halo of NGC~5907 were as contaminated as ours by ring
light and residual light from masked foreground stars.

\subsection{Comparison to the Results of Lequeux et al. and Rudy et al.}

The J, K, V and I surface brightness gradients in NGC~5907 are given in
\cite{Rudy97} for both their observations (J and K) and those of Lequeux et al.
(1996) (V and I). The additional B observations of \cite{Lequeux98}
do not have published numbers.  In comparing these data to our own, we first 
need to determine if the surface photometry was deep enough to detect the ring
around NGC~5907.  In the case of MBH, the answer is yes (Sec. 4.1).

Of the observations of \cite{Rudy97} and Lequeux et al. (1996), the evidence
as to whether the \cite{Rudy97} data did go faint enough to detect the ring in
the near IR is inconclusive.  Rudy et al. do remark that their surface
brightness profiles show an asymmetry, with the NE side of the galaxy having
more light at low levels than the SW side. However, neither we nor MBH see
this bump in our data.  In the case of Lequeux et al. (1996), it is apparent
both from the image and the surface brightness profiles they give that their
observations did not go faint enough to detect the ring.

We compare the perpendicular profiles along the minor axis of NGC~5907 in the 
8020$\rm \AA$ image to that given by Lequeux et al. (1996). for the I band, 
and by \cite{Rudy97} for the J and K band, in Figure~13, and tabulate these 
data in Table~7.  The zero point shift in the I band is 0.1, implying a zero 
point shift between ${\rm m_{6660}-m_{8020}}$ and $\rm R-I$ of about 0.2.  

There is generally good agreement among the J, K and 8020$\rm \AA$
perpendicular profiles within the quoted errors.  The one exception is the
presence of the above--mentioned ``bump'' on the NE side of the Rudy et al. 
J and K data that is not present in either our data, those of MBH (cf.
Figure~12, A cut) or those of Lequeux et al.  This indicates little detectable
color gradients in colors can be formed from these three passbands, although the
mutual errors are large enough to hide a significant color gradient, if it does
exist here.  The absence of quoted errors for the Lequeux et al. I band data
preclude an accurate assessment of agreement/disagreement. Even so, within the
cited errors of the other data sets, one finds differences between the Lequeux
et al. I band perpendicular profile and those of the other data sets only in
the inner NE 3 measured points.  In contrast, the J, K and 8020$\rm \AA$ data
for these three measured points are in very good agreement.  Once again,
exactly how each group has handled the foreground star contamination problem
likely leads to the differences seen in the inner part of the profile shown
in Figure~13.

\subsection{Colors of NGC~5907 Perpendicular to its Disk}

As is evident from the above comparison with other data, it is very difficult
to obtain reliable colors of galaxies at faint surface brightnesses.  The
errors of both passbands involved in forming the color quadratically combine,
and the resulting color differences are often small relative to this error. 
Such is clearly the case for the color gradients we can form perpendicular to
the disk of NGC~5907 in the $\rm m_{6660}-m_{8020}$ color we can form from our
data, as shown in Figure~14.  The error bars are formal 1-$\sigma$ errors per
measurement.  Since in these data 2-$\sigma$ to 3-$\sigma$ errors are likely
present as well, it is apparent that the color gradients 4 kpc and greater
from the disk of NGC~5907 are completely dominated by observational error.

Interestingly enough, within 4 kpc of the galaxy plane, we find that in many 
of the cuts the color gradients become bluer in $\rm m_{6660}-m_{8020}$ 
with increasing radius, with cut D2 being the notable exception.  In D2 alone,
the color gradient, on both sides of the galaxy, seems to get {\it redder} 
with distance from the midplane.

As can be seen also in Figure~13, colors formed outside the 4-kpc midplane 
region from any of the published passbands show little difference within the
mutual errors of observations (save for the questionable NE bump in J and K 
in the Rudy et al. data).

\subsection{The Limiting Factors in Faint Galaxy Surface Photometry}

We have already detailed the three sources of systematic error in
faint galaxy surface photometry --- systematic errors in determining the
flatfield, systematic errors in determining the sky background, residual
surface brightness from bright stars that one cannot easily mask.  Most 
observers try to take these factors into account (although not all explain
that they do in their papers; cf. \cite{Lequeux96}; \cite{Lequeux98}).  However,
there are two other sources of zero point errors in sky level determination
that are not discussed in the previous papers.  We include these now to 
complete our discussion of the data.

First is the fact discussed above that CCDs have wavelength-dependent quantum
efficiencies and this dependence is slightly different from pixel to pixel.
Galaxies and sky background can have very different colors, depending on which
part of the spectrum is observed, how bright the night sky lines are, how
bright the sky continuum is, etc.  If the difference in color is sufficient to
change the effective wavelength of a filter enough to change its quantum
efficiency in a detectable manner, the sky level estimated will be affected in
direct proportion to the percent change in QE. Take the R band observation as
an example, since the galaxy effective wavelength could be off by 50--100$\rm
\AA$ from the sky effective wavelength in the broad-band R, it is entirely
possible that the QE of the CCD could differ by 1 part in 1000 from galaxy
effective wavelength to sky effective wavelength. The zero point of the B, V
and I observations are similarly affected.

The issue for J and K observations are somewhat different, as there the sky
level is so bright that it is a triumph of instrumentation that one can get
a good estimate of sky down to less than 0.01\% accuracy.  Again, however,
if the IR Array QE is a function of effective wavelength, accuracy in estimate
of sky background in the broad band J and K filters will be limited by
the difference in QE between galaxy color and sky color.

We chose intermediate-band filters for the BATC survey for several reasons,
but one of the chief ones was to limit the bandpass sufficiently that
differences in color of the objects observed would not produce significant
differences in the effective wavelength of the filters.  In the case of the
6660$\rm \AA$ and 8020$\rm \AA$ filters, we designed them to avoid the brightest
(and most variable) night sky lines.  As such, we believe the change in QE
between galaxy and sky in our filters is much less than in the broad-band
filters.  A separate advantage, of course, is in a concomitantly much lower
sky background (12 times lower in the 6660$\rm \AA$ filter than MBH obtained 
at KPNO in the broad-band R).

The second source of non-modeled zero point error is the photometricity of the
telescope field of view and the inherent non-flatness of the night sky.  All
telescopes suffer from optical imperfections, most of which affect the
transmission of light from the sky as a function of distance from the center
of the field. It is primarily for this reason that the flatness of the night
sky has become the final arbiter of how well the CCD image has been
calibrated.  Yet, the night sky itself is not flat, which is pointed out
by Wild (1997).  In our degree-sized BATC CCD fields we clearly see the
gradient of the night sky getting brighter towards lower altitudes. With CCD
arrays getting to be degree-sized and greater, this problem is only enhanced. 
While one takes many separate sky images at different times of night --- so
the altitudinal gradients go in different directions on the CCD --- in the end
the cancellation of the gradients cannot be perfect.  Any residual gradient in
the sky will then be mistaken as a gradient in the sensitivity of the CCD
itself, and mistakenly divided out.  It should be subtracted out of the image,
as we have attempted to do in the present paper.

One advantage of using a Schmidt telescope for the BATC survey is that
we position the CCD in the center 1$^\circ$ of a focal plane that has
been specifically optimized to give as flat a field of view as possible
in the center of a 6$^\circ$ radius circle.  Such is true for any Schmidt
telescope, which is one reason they are noted for giving the most reliable
sky background estimates of any type of telescope.

The bottom line is that all sources of error conspire such that it is
extremely difficult to obtain sky background estimates and large-scale
uniformities to accuracies better than 0.1\% of the sky level. In defense of
our results, we can point to much lower sky level and narrower passbands.

\subsection{Does NGC~5907 Have a Halo?}

Two facts are clear from the above analysis:  1) Our data substantially agree
with the data in the literature for the surface brightness distribution seen
perpendicular to the disk of NGC~5907.  2) We find the light distribution
perpendicular to the disk of NGC~5907 to be highly {\it asymmetric} about the
galaxy midplane, with one side enhanced relative to the other in each MBH B, C
and D cut.

The difference between the present data set from those previously published
is a combination of the advantage of using the center field of a Schmidt
telescope as well as intermediate-band filters.  The use of the intermediate
band filters resulted in a much fainter sky; the large field of view of the
Schmidt permitted features such as the ring to be easily identified.
The large angular pixel size of our CCD (1.71$''$/pixel) is an advantage when
searching for faint surface brightness, as we minimize the number of pixels
per large surface area.  

Since we measure the ring around NGC~5907 to have an average surface
brightness of 28th $\rm 6660 \AA$ mag arcsec$^{-2}$, one can see from the
images themselves, Figures~2a and 6a, that NGC~5907 does not have a halo at
these faint surface brightnesses.  This fact is more clearly evident on the SW
side of the galaxy, the side away from the bulk of the ring.

We are forced to the conclusion that asymmetries in foreground star PSFs lead
to extra light around star masks, which combine with ring surface brightness
to produce what faint extensions of surface brightness we see in the MBH cuts.
This is due to the fact that the asymmetries in the surface brightnesses
derived from our images in the MBH cuts correlate with the degree of
foreground star contamination in four of the six non-centered cuts.  Overlap
with the ring similarly contributes to the asymmetries seen in the other
two of the non-centered cuts.  As we can see similar effects in the
star-masked image of MBH (their Figure 2), and as their data are in accord
with ours, we are equally forced to the conclusion that their detection of a
halo around NGC~5907 was an artifact caused by the same combination of ring
light and unaccounted faint wings from their foreground stars that affects our
data.

\section{CONCLUSION}

We have obtained very deep images of the edge-on galaxy NGC~5907 with two
intermediate-band filters on the BATC system (cf. Fan et al. 1996), $\rm
m_{6660}$ and $\rm m_{8020}$.  Via a detailed assessment of the sources of our
errors, we show that our limiting magnitudes (where the observational error 
reaches 1 mag arcsec$^{-2}$) are 29.0 mag arcsec$^{-2}$ in the 6660$\rm \AA$
image (corresponding to 28.7 in R-band measures) and 28.0 mag arcsec$^{-2}$ in
the 8020$\rm \AA$ image (close to the I-band system).  This is over one
magnitude fainter than previously published measurements, owing mostly to the
much darker sky as seen in our intermediate-band filters.

We use a new method of sky subtraction that both can account for most of the
foreground star contamination (by PSF-fitting the star profiles first, then
subtracting the fit), and permits reasonable interpolation of sky under the
galaxy region while also fitting the inherent lumpiness of the sky at low
surface brightnesses (using interactive fits of bidirectional, low-order
Legendre polynomials and heavy smoothing).  Our use of diffuser-smoothed dome
flats, already shown to give accurate photometric results over a wide range of
stellar colors (Fan et al. 1996) are again shown to give reliable flatfields
with these data.  This is critical, as it can be easily shown (cf. Wild 1997)
that the sky is not flat anywhere on degree size scales.

As first shown by Shang et al. (1998), we detail the evidence that NGC~5907
has an elliptical-shaped ring of emission around it, which mostly likely is
the remnant of a dwarf spheroidal galaxy that has been tidally torn apart.
Within errors of 0.3 -- 0.4 mag arcsec$^{-2}$, we show that the R--I colors of
the brighter part of this ring are consistent with the light coming from a
moderately metal-poor old stellar population.

We investigate the surface brightness distribution perpendicular to the disk 
of NGC~5907 by taking the same cuts through the galaxy as MBH.  We show that
all of the faint, extended surface brightness we observe perpendicular to the
disk of NGC~5907, and above an altitude of 5 kpc, is most likely due to
contamination by ring light and by residual effects from foreground star
contamination.  The possibility of a face-on warp existing in NGC~5907, 
suggested by our HI observations, is too confused with foreground star
contamination to permit us to prove or disprove its optical existence.

Our comparisons to the existing perpendicular profiles in the literature
for NGC~5907 find us in substantial agreement, once predicted zero point 
differences are removed, with these data.  Of particular note is that with
the known zero point difference between our 6660$\rm \AA$ filter and the
R band removed, the our data and those of MBH are in excellent agreement.

As such, we find that previous claims by Sackett et al. and others that
NGC~5907 has a faint halo to be an artifact due to a combination of two
factors.  One is the added light complications from the presence of a
previously-unknown ring around the galaxy.  The second factor is the
unaccounted residual effects of the foreground star masking procedures for
both our data and those of MBH at faint levels.  The lack of a halo around
NGC~5907 is evident directly from our deep images of this galaxy (Figures 2a
and 6a), once one understands that the ring we detect has an average 6660$\rm
\AA$ surface brightness of 28th mag arcsec$^{-2}$.

\acknowledgments

We thank Jinyao Hu and Xiaowei Liu for comments and discussions, Elias Brinks
for his help in obtaining the VLA data, Heather Morrsion for supplying us with
the data of MBH, and the referee for helpful comments.
The research done with the BATC Survey is supported by the Chinese Academy of
Sciences (CAS), the Chinese National Natural Science Foundation (CNNSF) and
the Chinese State Committee of Sciences and Technology (CSCST). It is also
supported in part by the U.S. National Science Foundation (NSF Grant
INT-93-01805), by Arizona State University, the University of Arizona and
Western Connecticut State University.

\clearpage

\begin{figure}
\figurenum{1}
\caption[fig01.ps]{The mean values of sky level, flux and FWHM for 20
comparison stars found in common on all the frames in 6660{\AA} filter. The
data of adjacent observations are displayed with different symbols. With these
parameters we selected out images with high signal-to-noise ratio so that we
can get data as accurate as possible. 
\label{fig1}} 
\end{figure}

\begin{figure}
\figurenum{2a}
\caption[fig02a.ps]{The 6660{\AA}, 26$^{\rm h}$20$^{\rm m}$ combined deep image of
NGC5907. The central square shows the area used in photometry, which is
1000$\times$1000 pixels. The ring around NGC~5907 can be seen faintly in this
image; its mean surface brightness is 28th $\rm 6660 \AA$ mag arcsec$^{-2}$.
\label{fig2a}}
\end{figure}

\begin{figure}
\figurenum{2b}
\caption[fig02b.ps]{The 8020{\AA}, 16$^{\rm h}$40$^{\rm m}$ combined image of 
NGC5907. Note the ring is of too faint a surface brightness to be seen in 
this image. 
\label{fig2b}}
\end{figure}

\begin{figure}
\figurenum{3}
\caption[fig03.ps]{Four examples of subtracting and masking stars and galaxies.
(a) A saturated bright star.  This PSF cannot be subtracted by DAOPHOT. Its
wing extends so far that the mask radius is larger than 130 pixels. Another
saturated star is found near this one. Their mask circles exclude most
contamination of their extended wings. (b) An unsaturated bright star.  The
residuals after the PSF is fit shows that the central pixels have large
fluctuations around the sky level and the residual wing of its PSF is still
partly visible.  The mask radius used is 21 pixels. (c)  A star of moderate
brightness. The residuals after PSF is fit show significantly smaller
variation in its center than for the bright star, and the PSF wings for this
star are smaller as well.  The mask radius used is 10 pixels. (d) A faint
galaxy. The PSF for a galaxy is much gradual than that of stars, such that it
can not be subtracted properly by DAOPHOT using a stellar PSF.  The mask
radius used is 9 pixels, while a star with the same brightness needs a mask
radius of only 5 pixels or less. \label{fig3}}
\end{figure}

\begin{figure}
\figurenum{4}
\caption[fig04.ps]{The final mask that will be applied to the
6660{\AA} image to account for foreground stars and background galaxies and a
circle region around NGC~5907. The central circle indicates the region in which 
all pixels around NGC~5907 are masked; the curved area around this
circle masks the ring we found around this galaxy.  Separately, for profile 
determinations the dust lane of NGC~5907 is also masked.  The mask radii of 
most faint stars are too small in size and low in contrast to be clearly 
displayed in this image. \label{fig4}}
\end{figure}

\begin{figure}
\figurenum{5}
\caption[fig05.ps]{The four steps of fitting background. (a) Fit the image 
row by row with Legendre polynomials of order 3 or less.
(b) Fit the image column by column in the same way. (c) Take the average of
(a) and (b). (d) Smooth (c) with a Gaussian function.  It can be found
that the final fitted background is too complicated to be expressed
by any 2-D analytical function.  The net results of the fit underneat the
galaxy region are shown in the slices in Figures 7 and 8.\label{fig5}}
\end{figure}

\begin{figure}
\figurenum{6a}
\caption[fig06a.ps]{The background--subtracted image in 6660{\AA}. The ring
around the galaxy is very clear. It is a near-perfect, but not complete
ellipse, with a major axis of $13.7'$, minor axis of $7.2'$ and eccentricity
$e=0.84$ (cf. Shang et al. 1998). The major axis size is 44 kpc at a predicted
distance of 11 Mpc (assuming H$_0=75$ km s$^{-1}$ Mpc$^{-1}$). The area within
the rectangle marked in the image was measured to determine the
average surface brightness of the brightest area of the ring in 6660{\AA} and
8020{\AA}, so that we can get color index of the ring with reasonable accuracy.
A schematic figure is inserted in bottom-right corner of the image to show the
relative position of the ring and the galaxy. The solid lines with labels
indicate the cuts along which surface photometry of the galaxy halo is done.
These cuts are the same as that of Sackett et al. (1994) and MBH. The dashed 
lines show the centers of the slices made to check effect of 
background-subtraction.  Note the extent to which foreground stars exist to 
affect the surface brightness distributions we can obtain in the MBH cuts. 
\label{fig6a}}
\end{figure}

\begin{figure}
\figurenum{6b}
\caption[fig06b.ps]{The background-subtracted image in 8020{\AA}. \label{fig6b}}
\end{figure}

\begin{figure}
\figurenum{7}
\caption[fig07.ps]{Four slices in the background--subtracted 6660{\AA} image to
show the accuracy of background fitting and local surface brightness of the
ring feature. The positions of the center of each slice relative to the galaxy
and ring are repreented in Fig. 6a by the dashed lines.
Each slice is averaged over 31 lines of the image, and only non-masked pixels
are averaged. In the case that all 31 points are masked, a zero value is
assigned. Also shown are a straight line indicating the zero level.  Note also
the small, but still apparent, residual effect of our star subtraction/masking
procedure, a product of the compromise made in the star masking procedure. 
The lower part of each panel (i.e., below the zero lines) shows the residuals
of subtracted sky background for each slice.  The upper line in each panel
shows the residuals produced by our sky fitting procedure, while the lower
line shows that produced by normal two-dimensional function fitting. They are
shifted down by 110 ADU and 150 ADU seperately for ease of display. One can
compare them and easily find that our procedure produce smaller residuals. The
arrows in the three panels show the positions where the slices cross the ring.
\label{fig7}}
\end{figure}

\begin{figure}
\figurenum{8}
\caption[fig08.ps]{The slices in the background--subtracted 8020{\AA} image,
shown in the same format as in Figure~7.
\label{fig8}}
\end{figure}

\begin{figure}
\figurenum{9}
\caption[fig09.ps]{The stellar warp of NGC5907 measured in two bands. The
circles are from 6660{\AA} image and the squares are from 8020{\AA} image. The
x axis is along the galaxy plane, which is determined by fitting the symmetry
axis of the galaxy. The stellar warp is obvious even in the cut nearest the
galaxy center. \label{fig9}} 
\end{figure}

\begin{figure}
\figurenum{10}
\caption[fig10.ps]{The surface brightness profiles in the 6660{\AA} passband
for the seven MBH--consistent cuts perpendicular to the galaxy plane. Each
panel corresponds to one cut shown in Fig. 6a and the separate measurements of
the two sides of the galaxy plane are shown. The data of NE and SW side of the
galaxy plane are denoted by squares and triangles, respectively. The open
symbols denote the points affected by the ring. Central regions of the galaxy
are excluded due to star and dust masking. The asymmetries at faint surface
brightnesses that exist in these cuts are discussed in the text.
\label{fig10}}
\end{figure}

\begin{figure}
\figurenum{11}
\caption[fig11.ps]{The surface brightness profiles in 8020{\AA}. \label{fig11}}
\end{figure}

\begin{figure}
\figurenum{12}
\caption[fig12.ps]{Comparison of our 6660{\AA} data with that of MBH, with a
zero point shift of 0.3 mag applied to the MBH data to account for the
different passbands used. The circles are the data of MBH that were sent to us
by Dr. Morrison, which go fainter than those published by MBH in their actual
paper. The solid circles are the reliable data claimed by them and published
in Sackett et al. 1994, while the open circles (fainter than 27 mag
arcsec$^{-2}$) were not published. The other symbols have the same meaning as
Fig.10.  Note the general good agreement between our data and those of MBH
(save for cuts A and C2) even at surface brightnesses fainter than 27th mag
arcsec$^{-2}$.
\label{fig12}}
\end{figure}

\begin{figure}
\figurenum{13}
\caption[fig13.ps]{Comparison of our 8020{\AA} data with infrared data from 
Rudy et al. (J and K) and Lequeux et al. (I), with zero point applied to
the published data as discussed in the text.  Note the general good agreement
among the different passbands, with their mutual errors, except for the
``bump'' seen in the data of Rudy et al. but not in the other data sets.
\label{fig13}}
\end{figure}

\begin{figure}
\figurenum{14}
\caption[fig14.ps]{The m$_{6660}$-m$_{8020}$ color profiles of NGC5907.
Measurable color gradients become bluer with distance from the galaxy midplane
in all cuts except D2, where the color gradient actually becomes redder with
distance from the midplane. The meaning of the symbols is the same as they in
Fig. 10 and Fig. 11.  It is apparent that the errors in the colors preclude any
meaningful measurement of color gradient outside of 4 kpc from the plane of
this galaxy. \label{fig14}}
\end{figure}


\clearpage

\begin{thebibliography}{}

\bibitem[Barnaby \& Thronson 1992]{Barnaby92}
Barnaby, D. \& Thronson, H. A. 1992, AJ, 103, 41

\bibitem[Chen et al. (1999)]{Chen99}
Chen, J.-S. et al. 1999, in preparation

\bibitem[James \& Casali (1996)]{James96}
James, P. \& Casali, M. M. 1996, Spectrum, March, 14

\bibitem[Fan et al. 1996]{Fan96}
Fan, X.-H. et al. 1996, AJ, 112, 628

\bibitem[Lasker et al. 1990]{Lasker90}
Lasker, B. M., et al., 1990, AJ, 99, 2019

\bibitem[Lequeux et al. 1996]{Lequeux96}
Lequeux, J., Fort, B., Dantel-Fort, M., Cuillandre, J.-C. \& Mellier, Y.
1996, A\&A, 312, L1

\bibitem[1998]{Lequeux98}
Lequeux, J. et al. 1998, astro-ph/9804109

\bibitem[Morrison et al. (1994)]{Morrison94}
Morrison, H. L., Boroson, T. A., \& Harding, P. 1994, AJ, 108, 1191 [MBH]

\bibitem[Oke--Gunn (1983)]{Oke83}
Oke, J. B. \& Gunn, J. E., 1983, ApJ, 266, 713

\bibitem[Peterson et al. 1993]{Peterson93}  
Peterson, R. C., Dalle Ore, C. M. \& Kurucz, R. L. 1993, ApJ, 404, 333

\bibitem[Rudy et al. (1997)]{Rudy97}
Rudy, R. J., Woodward. C. E., Hodge, T., Fairfield, S. W. \&
Harker, D. E. 1997, Nature, 387, 159

\bibitem[Sackett et al. (1994)]{Sackett94}
Sackett, P. D., Morrison, H. L., Harding, P., \& Boroson, T. A.
1994, Nature, 370, 441

\bibitem[Sancisi 1976]{Sancisi76}
Sancisi, R. 1976, A\&A, 53, 159

\bibitem[Sancisi (1993)]{Sancisi93}
Sancisi, R. 1993, private communication

\bibitem[Sasaki 1987]{Sasaki87}
Sasaki, T. 1987, PASJ, 39, 849

\bibitem[Shang et al. (1998)]{Shang98}
Shang, Z.-H. et al. 1998, ApJL, accepted

\bibitem[Skurtskie et al. 1985]{Skurtskie85}
Skurtskie, M. F., Shure, M. A. \& Beckwith, S. 1985, ApJ, 299, 303

\bibitem[Stetson 1987]{Stetson87}
Stetson, P. B. 1987, PASP, 99, 191

\bibitem[Stetson 1990]{Stetson90}
Stetson, P. B. 1990, PASP, 102, 932

\bibitem[Tonry \& Schneider (1988)]{Tonry88}
Tonry, J. \& Schneider, D. P. 1988, AJ, 96, 807

\bibitem[Tonry et al. 1990]{Tonry90}
Tonry, J. L., Ajhar, E. A., \& Luppino, G. A. 1990, AJ, 100, 1416

\bibitem[van der Kruit (1979)]{van79}
van der Kruit, P. C. 1979, A\&AS, 38, 15

\bibitem[van der Kruit \& Searle 1981]{van81}
van der Kruit, P. C. \& Searle, L., 1981, A\&AS, 95, 105

\bibitem[Wild 1997]{Wild97}
Wild, W. J. 1997, PASP, 109, 1269

\end{thebibliography}
\end{document}




\newpage

\begin{deluxetable}{ccccccc}
\tablenum{3}
\tablecolumns{7}
\tablewidth{0pt}
\tablecaption{Parameters of the combined images \label{table3}}
\tablehead{
\colhead{Filter} &  \colhead{Total} & \colhead{Seeing\tablenotemark{a}} 
&  \colhead{Sky level} 
& \colhead{$\sigma_{\rm sky}$\tablenotemark{b}}
&  \colhead{Readout noise\tablenotemark{c}} 
& \colhead{Zero point\tablenotemark{d}} \\
\colhead{} & \colhead{exposure times} & \colhead{\footnotesize FWHM(pixel)} 
& \colhead{({\footnotesize mag/sq arcsec})} 
 & \colhead{({\footnotesize ADU})} & \colhead{(e$^-$)}  
& \colhead{(1{\footnotesize ADU}/s)}} 
\tablenotetext{a}{Mean FWHM of all the stars in the field.}
\tablenotetext{b}{The standard error of sky background per pixel.}
\tablenotetext{c}{Total readout noise introduced into the final combined images
per pixel.}
\tablenotetext{d}{Photometry zero point calibrated by measuring Oke--Gunn 
standard stars.}
\startdata
6660$\rm \AA$ & $\rm 26^h10^m$  & 2.3   &  21.26    & 105   & 110   & 19.08  \nl
8020$\rm \AA$ & $\rm 16^h40^m$  & 2.5   &  19.91    & \phn 85   & \phn 85   & 17.79  \nl
\enddata
\end{deluxetable}

\newpage

\begin{deluxetable}{lll}
\tablenum{4}
\tablecolumns{3}
\tablewidth{0pt}
\tablecaption{Summary of limiting error sources in a 45$\times$45 pixel bin
in the 6660{\AA} image \label{table4}} 
\tablehead{
\colhead{Source of variation} & \colhead{ADU} & \colhead{\phn \%} }
\startdata
\multicolumn{3}{c}{\it Reducible noise}\nl
Readout noise &   \phn 0.6  &   \phn 0.002 \nl
Photon statistics &  \phn  2.1  &  \phn  0.006 \nl
Bias and Dark &   \phn  0.02   &  \phn   0.00005   \nl
Flatfielding errors  & \phn  0.08 & \phn  0.0002\nl
Surface brightness fluctuations & \phn  0.3 & \phn  0.0008\nl
\nl
\multicolumn{3}{c}{\it Irreducible noise}\nl
Large-scale FF errors & \phn  3.7 & \phn  0.01 \nl
Systematic error of background subtraction &  18.0 & \phn  0.05 \nl \nl
\tableline
\nl
Total &  23.9 &  \phn 0.06 \nl 
\enddata       
\end{deluxetable}%
